\documentclass[doublecol,figures]{epl2}

\title{Deviation from Gaussianity in the cosmic microwave background
       temperature f\/luctuations}
\shorttitle{Nonextensivity of the CMB}

\author{A. Bernui\inst{1} \and C. Tsallis\inst{2} \and T. Villela\inst{1}}
\shortauthor{A. Bernui \and C. Tsallis \and T. Villela}  

\institute{
\inst{1} Instituto Nacional de Pesquisas Espaciais, Divis\~{a}o de Astrof\'{\i}sica \\ 
         Av. dos Astronautas 1758, 12227-010 S\~ao Jos\'e dos Campos, SP, Brazil
\\
\inst{2} Centro Brasileiro de Pesquisas F\'{\i}sicas \\ 
         Rua Xavier Sigaud 150, 22290-180 Rio de Janeiro, RJ, Brazil
}

\pacs{98.80.Es}{Observational cosmology}
\pacs{98.70.Vc}{Background Radiations: statistical properties}
\pacs{05.90.+m}{Statistical Physics, thermodynamics, and nonlinear dynamical systems}

\abstract{
Recent measurements of the temperature f\/luctuations of the Cosmic Microwave 
Background (CMB) radiation from the WMAP satellite provide indication of a 
non-Gaussian behavior. 
Although the observed feature is small, it is detectable and analyzable.
Indeed, the temperature distribution $P^{\mbox{\sc cmb}}(\Delta T)$ 
of these data can be quite well f\/itted by the anomalous probability 
distribution emerging within nonextensive statistical mechanics, based on 
the entropy
$S_q \equiv k \{ 1-\int \, \upd x \, [P(x)]^q \} /(q\!-\!1)$ 
($S_1 \!=\! - k \int \, \upd x \, P(x) \, \mbox{\rm ln}[P(x)]$).
For the CMB frequencies analysed, $\nu=$ 40.7, 60.8, and 93.5 GHz, 
$P^{\mbox{\sc cmb}}(\Delta T)$ is well described by
$P_q(\Delta T) \propto 1/[1+(q-1) B(\nu) (\Delta T)^{\,2}]^{1/(q-1)}$, with
$q = 1.04 \pm 0.01$, the strongest non-Gaussian contribution coming from the 
South-East sector of the celestial sphere. 
Moreover, Monte Carlo simulations exclude, at the 99\% conf\/idence level,
$P_1(\Delta T) \propto e^{- B(\nu) (\Delta T){\,^2}}$ 
to f\/it the three-year data.}

\begin{document}

\maketitle


\section{Statistical properties of the CMB temperature f\/luctuations}
Since the early 90's a set of precise observations showed that the angular 
distribution of the Cosmic Microwave Background (CMB) radiation presents 
tiny f\/luctuations, of the order of $10^{-5}$, around the equilibrium 
temperature $T_0$.
These CMB temperature f\/luctuations, denoted $\Delta T(\theta,\phi)$,
are functions of the direction of observation in the celestial sphere
$(\theta,\phi)$ and can be expressed by using a spherical harmonic expansion
$\Delta T(\theta,\phi)/T_0 = \sum_{\ell \, m} \, a_{\ell \, m} Y_{\ell \, m}(\theta,\phi)$. 
Standard cosmological theory predicts that the complex $a_{\ell \, m}$ 
modes are statistically isotropic and Gaussian random f\/ields of zero mean
($\langle a_{\ell\,m} \rangle = 0$) and non-zero variance
($\langle a^{\ast}_{\ell'\,m'} \, a^{}_{\ell\,m} \rangle \neq 0$).
The fact that $\Delta T$ is a real quantity yields consistency 
relationships between the real and imaginary parts of $a_{\ell\,m}$ and 
$a_{\ell\,-m}$.
{\em Gaussian randomness} means that both real and imaginary parts of 
the $a_{\ell\,m}$ are each an independent random variable obeying a
Gaussian distribution of zero mean.
{\em Statistical isotropy} means that the variance, which in principle
could be different for each $\ell$ and $m$, depends only on $\ell$, i.e.
$\langle a^{\ast}_{\ell'\,m'} \, a^{}_{\ell\,m} \rangle
= {\cal C}_{\ell} \, \delta_{\ell'\,\ell} \, \delta_{m'\,m}$.
Thus, in the standard inf\/lationary big-bang model, one expects that
the CMB temperature f\/luctuations $\Delta T$ should be appropriately
described by a Gaussian distribution.

The recent release of the three-year high angular resolution and low 
instrument noise CMB measurements from the Wilkinson Microwave Anisotropy 
Probe (WMAP)~\cite{WMAP3y} offered the opportunity to analyse, with 
unprecedent precision, their temperature distribution in order to test 
the Gaussian hypothesis. 
These WMAP maps~\cite{Lambda}, with parameter resolution $N_{side}=512$, 
divided the celestial sphere into $3,\/145,\/728$ equal sized areas, termed 
pixels, each one with an assigned highly precise $\Delta T$ value. 
Because measurements in the region around the Galactic Plane are strongly
contaminated with foregrounds, they shall be eliminated from the data set 
in analysis by applying a cut-sky mask.
The remaining $\sim 2.5$ million pixels guarantee a f\/ine adjustment of the 
temperature distribution histogram (also termed {\it one-point distribution 
function}) to a smooth curve, thus minimizing the uncertainty of any 
(Gaussian or non-Gaussian) f\/itting.

Through the Minkowski functional's analysis, the WMAP team placed tight new
limits on non-Gaussian temperature f\/luctuations in the CMB, thus concluding
that these data are consistent with Gaussianity~\cite{Komatsu,Analysis}.
However, several kinds of non-Gaussian features have been reported in the 
f\/irst-year and three-year WMAP data~\cite{NonGaussian}.
Clearly, the detailed study of the Gaussian random hypothesis must take
into account the possibility that deviations from Gaussianity may have
non-cosmological origins, such as unsubtracted foreground contamination,
instrumental noise, and/or systematic ef\/fects~\cite{Foregrounds}.
Nonetheless, it is also possible that deviations from Gaussianity, although
small, may be intrinsic to the CMB radiation~\cite{BTV} or have some other
cosmological origin such as cosmic strings~\cite{JS}. 
The related issue regarding the violation of the large-scale statistical 
isotropy, termed the North-South asymmetry, has also been extensively 
investigated in a variety of analyses, using dif\/ferent 
mathematical-statistical tools, with the f\/irst-year WMAP data~\cite{Copi1} 
and more recently also with the three-year data~\cite{Copi3}.

Given the potential importance of these reported non-Gaussian features for our 
understanding of the physics of the early universe, all these analyses must be 
considered as non-exhaustive and further studies have to be performed.
Although these non-Gaussian f\/indings~\cite{BTV} -- obtained with the 
f\/irst-year WMAP data -- could be interpreted as unremoved foregrounds 
and/or instrumental noise, a comparative analysis with three-year data, 
where the signal to noise ratio increases due to a greater number of 
observations {\it per} pixel, could conf\/irm or discard some of these 
possible explanations, or simply put better limits of their effect on the 
CMB data.
This converts the analyses of the statistical properties of the WMAP 
three-year data into a necessary and important task. 

Nonextensive statistical mechanics appears to be suitable to describe 
statistical properties of long-range correlated systems~\cite{CarusoTsallis}, 
like the primordial CMB radiation filling the universe. 
For such systems, correlations are expected to break down the Gaussian, 
i.e., essentially uncorrelated, properties~\cite{hamity}. 
In that situation, the $q$-index can be seen as a scalar measure of the 
correlation range present in the system.
For example, for an attractive two-body potential decaying as 
$1/r^{\alpha}, \alpha > 0$, in a $d$-dimensional classical system, we may 
 expect $q$ to increase above unity when $\alpha/d$ decreases below 
unity~\cite{rapisardapluchino}. 

In what follows we perform a brief overview of the non-Gaussian 
distributions emerging from nonextensive statistical mechanics.

\section{Basics of nonextensive statistical mechanics}
The probability distribution function $P_q$ in the frame of the nonextensive 
statistical mechanics results~\cite{BTV} from the optimization of the 
$q$-entropy def\/ined~\cite{Tsallis} (see~\cite{aplic} for various 
applications, and~\cite{douglas} for a new experimental conf\/irmation) 
as follows ($q \in \Re$) 
\begin{eqnarray} \label{eq.1}
S_q \equiv k \, \frac{1-\int \upd x \, [P(x)]^q }{q-1}  \, , \,\, \\
S_1 = S_{BG} \equiv -k \int \upd x \, P(x)\,\mbox{\rm ln}[P(x)] \, , \nonumber
\end{eqnarray}
satisfying suitable constraints ($BG$ stands for {\it Boltzmann-Gibbs}).
Given two independent systems $\Sigma_A, \Sigma_B$ (in the sense that 
$P^{\Sigma_A \cup \Sigma_B}_{q} = P^{\Sigma_A}_q \, P^{\Sigma_B}_q$), 
the pseudoadditivity property says that 
$S^{\Sigma_A \cup \Sigma_B}_q 
= S^{\Sigma_A}_q + S^{\Sigma_B}_q + (1-q) S^{\Sigma_A}_q S^{\Sigma_B}_q$. 
In other words, $(1-q)$ is a measure of the nonadditivity
of the total system $\Sigma_A \cup \Sigma_B$. 
Clearly, the standard, extensive, statistical mechanics is recovered in the 
$q \rightarrow 1$ limit.

\noindent
From Eq.~(\ref{eq.1}) one deduces~\cite{BTV} (see also~\cite{Tsallis-CLT} 
regarding the connection with the $q$-Central Limit Theorem)
\begin{equation} \label{P-q}
P_q(\Delta T) 
= A_q \, e_q^{-B_q {(\Delta T})^2} \, ,
\end{equation}
where in general $B_q$ depends on the $q$-index, and $A_q$ is the 
normalization constant obtained in such a way that $\int\, \upd x \,P_q(x)=1$.
The $q$-exponential is def\/ined by 
\begin{equation} \label{q-exp}
e_q^z \equiv [1 + (1-q)z]^{1/(1-q)} \, , \,\,\,\,\,\,
\mbox{\rm for $[1 + (1-q)z] \geq 0$} \, ,
\end{equation}
and $e_q^z = 0$ otherwise.
In the limit $q \to 1$, we recover the Gaussian distribution
\begin{equation} \label{P-Gauss}
P_q(\Delta T) \to P_1 \equiv P^{\mbox{\rm\footnotesize Gauss}}(\Delta T) 
= A\, e^{-B(\nu) {(\Delta T})^2} \, ,
\end{equation}
where
$A \equiv 1/(\sigma_{\nu}\sqrt{2\pi}), \, B(\nu) \equiv 1/(2\,\sigma_{\nu}^2)$, 
and $\sigma_{\nu}^2$ is the variance of the Gaussian distribution.

We shall use the nonextensive probability distribution $P_q$, def\/ined in 
Eqs.~(\ref{P-q}) and~(\ref{q-exp}), to study the CMB temperature 
distribution $P^{\mbox{\sc cmb}}$ from the WMAP three-year data. 
Notice that by plotting the temperature distribution histogram in the form of 
the Number of Pixels {\it versus} $(\Delta T/\sigma_{\nu})^2$ any 
Gaussian distribution f\/its a straight line, thus recognizing departures 
from Gaussianity (i.e., from linearity) becomes easy.
For a clear comparison between the Gaussian (linear) and the 
nonextensive distributions we shall plot both distributions $P_q$ and 
$P^{\mbox{\rm\footnotesize Gauss}}$ having equal value at the initial point 
$\Delta T=0$ (that is, $A_q = A = 1$). 
Moreover, we observe that the best-f\/itting tangent line at that point has 
equal slope for both distributions, which implies that $B_q = B(\nu)$ (this 
procedure is consistent with the fact that $d\,e_q^z/dz|_{z=0}=1,\,\forall q$). 
Therefore, the best-f\/itting of the CMB data through the nonextensive
distributions is obtained by adjusting only the $q$-index.

\section{WMAP data analyses}
We analyse the WMAP three-year data by exploring their temperature 
distribution in terms of Gaussian and nonextensive probability distributions.
Since any non-Gaussian signature is revealed as non-linearities when the 
temperature distributions are plotted in the form of the Number of Pixels 
{\it versus} $(\Delta T/\sigma_{\nu})^2$, we present our results using this 
type of plots. 
We use the $\chi^2$/degree of freedom (dof) estimator test to f\/it the data
with the best Gaussian and nonextensive distributions functions.

We investigate these data corresponding to the {\it Q}, {\it V}, and 
{\it W}-bands (central frequencies at 40.7, 60.8, and 93.5 GHz, respectively) 
by analysing the eight foreground reduced differencing assemblies (DAs) 
available at~\cite{Lambda}, 
namely {\it Q}1, {\it Q}2, {\it V}1, {\it V}2, {\it W}1, {\it W}2, {\it W}3, 
and {\it W}4. 
These DA CMB maps were corrected by the WMAP team for the
Galactic foregrounds (free-free, synchrotron, and dust emission) using the
3-band, 5-parameter template f\/itting method described in~\cite{WMAP3y}.
However, the foreground removal is only applicable to regions outside
the Kp2 mask (which cuts $\sim 15\%$ of the sky data).
The more severe cut, represented by the Kp0 mask (which excludes $\sim 23\%$ 
of the sky data), is recommended~\cite{WMAP3y} for the statistical 
analysis of the CMB maps. 
Thus, for our analyses we have used the Kp0 mask.
For the sake of completeness, we have also examined the foreground-reduced 
coadded maps {\it Q} (= {\it Q}1$\oplus${\it Q}2), {\it V} (= {\it V}1$\oplus${\it V}2), 
and {\it W} (= {\it W}1$\oplus${\it W}2$\oplus${\it W}3$\oplus${\it W}4), as 
well as the coadded {\it V}$\oplus${\it W} map, formed by the six DAs 
{\it V}1$\oplus${\it V}2$\oplus${\it W}1$\oplus${\it W}2$\oplus${\it W}3$\oplus${\it W}4, 
and the coadded {\it Q}$\oplus${\it V}$\oplus${\it W} map, formed by the eight 
DAs (the $\oplus$ symbol means the coadded sum of CMB maps as def\/ined 
in~\cite{Jarosik}).

Finally, we also investigated the temperature distribution in the four 
quadrants of the celestial sphere (NE, NW, SW, and SE) in order to test the 
claimed large-scale anomaly~\cite{Copi1,Copi3} in the CMB temperature 
f\/luctuations data.

As pointed out in~\cite{JS}, the signal measured at any pixel in the 
microwave sky is made of several components 
$T_{\mbox{\rm\small pixel}} = T_{\mbox{\rm\small foregrounds}} 
+ T_{\mbox{\rm\small noise}} + T_{\mbox{\sc cmb}}$, 
corresponding to foreground contamination signals, the noise from the 
instruments, and the CMB temperature f\/luctuations, respectively. 
Foreground contributions are expected around the Galactic Plane and from
point sources. 
Given the fact that the instrument noise in the map is inhomogeneous 
(higher in the Ecliptic Plane than near the poles), the contribution from 
$T_{\mbox{\rm\small noise}}$ is important.
For the subtraction of the foregrounds from CMB maps we used, as mentioned 
above, the most severe cut-sky, i.e., the Kp0 mask (for comparison, we also 
made the analyses with the Kp2 mask). 
The masks are provided by the WMAP team~\cite{WMAP3y} and allow for the 
selective exclusion of portions of the sky basically due to contaminating 
radiative processes from our galaxy, including a 0.6 degree radius exclusion 
area around known point sources in the celestial sphere.
The application of a mask to a CMB map simply means that, for a given pixel,
a mask value of zero implies that it is excluded; a value of one means that 
it is accepted.
Moreover, since the Galactic foregrounds contribute as positive 
temperature contaminations, we perform the analyses of the one-point 
distributions considering only the negative temperature f\/luctuations data.
In this way, the f\/irst term in the above equation is dropped and one is 
left with CMB signal plus the signal from the instrument noise (the signal 
noise is Gaussian {\it per} observation~\cite{JS,Jarosik}).
What converts the instrument noise in a non-Gaussian signal is the 
non-uniform way in which the CMB sky was observed, because some regions 
were largely more observed than others, making such a noise to be dif\/ferent 
from pixel to pixel.
The way out to deal with this problem is to characterize its shape and 
intensity through Monte Carlo realizations of similar sky data observations.

Consider the remaining ef\/fects by def\/ining~\cite{JS} the variance of the 
total signal $T_{\mbox{\rm\small pixel}}$ as 
\begin{eqnarray} \label{sigmaG}
\sigma_{\nu}^2 \equiv \sigma^2_{\mbox{\rm\small pixel}} =
\frac{\sigma_0^2}{n_{i}} + \sigma_{_{\mbox{\sc cmb}}}^2 \, ,
\end{eqnarray}
where $n_i$ is the number of observations for the $i$\/th pixel, $\sigma_0^2$ 
is the noise variance {\it per} observation of a given instrument~\cite{WMAP3y}, 
and $\sigma_{\mbox{\sc cmb}}^2$ is the variance of the CMB temperature
f\/luctuations.
The mean contribution of the instrumental noise can be estimated by
considering the ef\/fect of the dif\/ferent number of observations for each
pixel. 
This can be done with the ef\/fective variance due to the non-stationary 
instrument noise~\cite{JS} 
\begin{eqnarray} \label{sigmaNoise}
\sigma^2_{\mbox{\rm\small noise}}(\nu) = \frac{\sigma_0^2 \sum^N_{i=1} (1/n_i)}{N}\, ,
\end{eqnarray}
where  $N$ is the total number of pixels considered in the analysis of the 
map. 
The values of $\sum (1/n_i)$ for each WMAP DA map are given in 
table~\ref{table1} (Kp0 mask case) 
The total number of pixels $N$ with CMB signal corrected for Galactic 
contaminations results from the application of the cut-sky mask, that is 
$N=N_{\mbox{\rm\footnotesize Kp0}}=2,\/407,\/737$ for the Kp0 mask, or 
$N=N_{\mbox{\rm\footnotesize Kp2}}=2,\/664,\/479$ for the Kp2 mask. 
Thus, the ef\/fective noise variances $\sigma^2_{\mbox{\rm\small noise}}(\nu)$ 
and the variances $\sigma_{\nu}^2$ leads to the CMB variance 
\begin{eqnarray} \label{sigmaCMB}
\sigma^2_{_{\mbox{\sc cmb}}} = \sigma_{\nu}^2 
- \sigma^2_{\mbox{\rm\small noise}}(\nu)\,.
\end{eqnarray}
Although both $\sigma_{\nu}^2$ and $\sigma^2_{\mbox{\rm\small noise}}(\nu)$ 
depend on the map under analysis, their dif\/ference is independent of 
the map.
In other words, if the treatment of instrumental noise is correct, the CMB 
variance $\sigma^2_{\mbox{\sc cmb}}$ should be the same for the maps under 
investigation ({\it Q}, {\it V}, and {\it W}).

The present best-f\/it data analyses were obtained according to the 
$\chi^2$/degree of freedom (dof) estimator test. 
Thus, the typical $\chi^2 / 200$ values for the Gaussian f\/its are 
$2 \times 10^{-2}$, and the $\chi^2 / 200$ values for the nonextensive 
temperature distribution are $2 \times 10^{-3}$.
Analyses performed using 300 dof and 400 dof, instead of 200 dof, result in 
similar best-f\/itting values. 
In addition, we have not detected any significative difference using the Kp2 
mask instead of the Kp0 mask.

\begin{figure*} [t] 
\vspace{-1cm}
\onefigure[width=18cm,height=23cm]{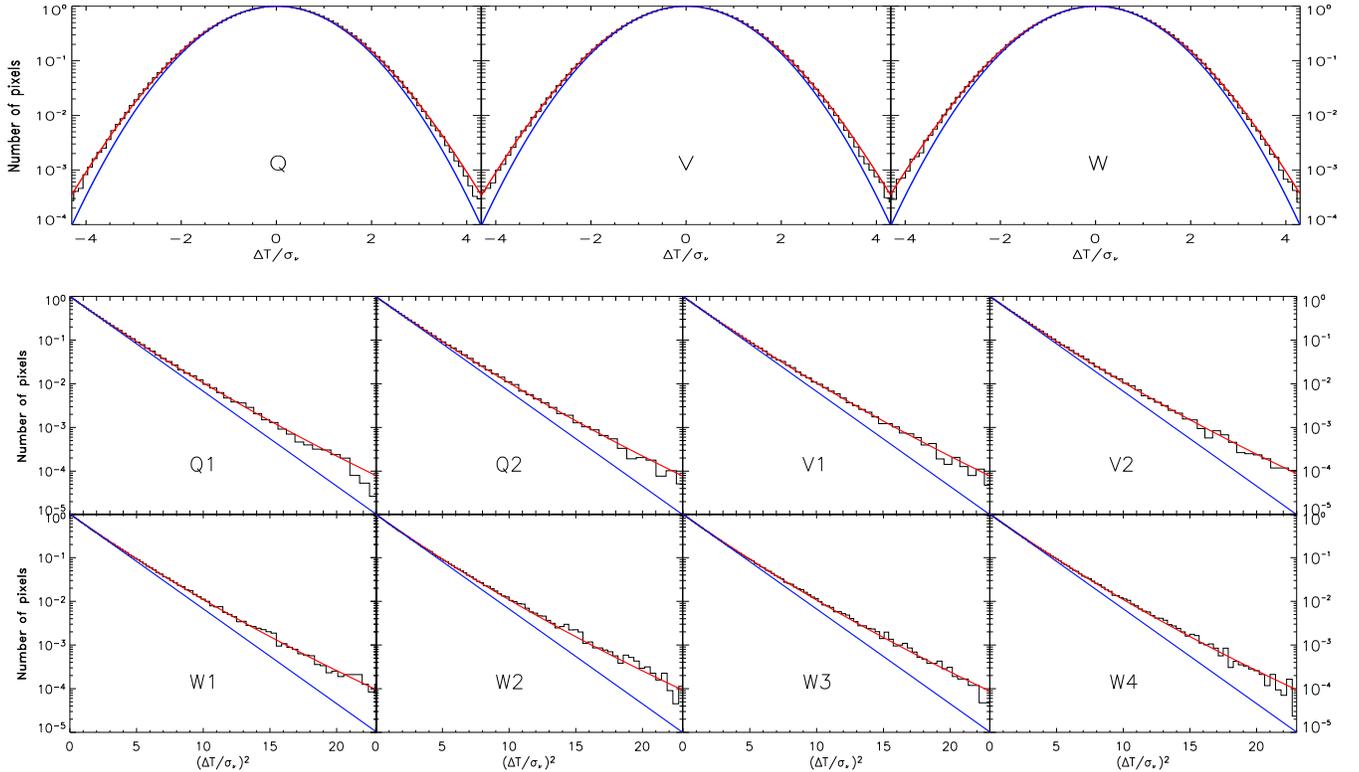}
\vspace{-11.cm}
\caption{
{\bf Top:} 
Fits to the (positive and negative) WMAP CMB temperature f\/luctuations data,  
corresponding to the {\it Q}, {\it V}, and {\it W} coadded maps  (after the Kp0 
cut-sky), in the  
{\sc number of pixels} {\it versus} $\Delta T / \sigma_{\nu}$ plots.
We show the $\chi^2$ best-f\/its: Gaussian distribution (blue curve) with 
$\sigma_{{\mbox{\sc q}}}=104\, \mu$K, $\sigma_{{\mbox{\sc v}}}=118\, \mu$K,
and  $\sigma_{{\mbox{\sc w}}}=131\, \mu$K, respectively, 
and each nonextensive distribution (red curve) $P_q$ with $q=1.04$.
{\bf Bottom:}
Similar analysis, but now with the eight DA WMAP maps ({\it Q1},...,{\it W4}) 
after applying the Kp0 mask.
We plotted the {\sc number of pixels} {\it versus} 
$(\Delta T/\sigma_{\nu})^2$ to enhance the non-Gaussian behavior.
To avoid possible unremoved Galactic foregrounds, we consider only the 
negative temperature f\/luctuations.
Again, we show the $\chi^2$ best-f\/its: Gaussian distribution (blue curve) 
and nonextensive distribution (red curve) $P_q$, now with $q=1.04 \pm 0.01$.}
\label{figure1}
\end{figure*}

Our analyses of the eight DAs is as follows.
Once the variances $\sigma_{\nu}^2$ have been determined through the $\chi^2$ 
best-f\/it Gaussian temperature distribution (blue lines in fig.~\ref{figure1}), 
we use the ef\/fective noise variance $\sigma^2_{\mbox{\rm\small noise}}$, 
given in Eq.~(\ref{sigmaNoise}), to calculate the CMB variances 
$\sigma^2_{\mbox{\sc cmb}}$ (see table~\ref{table1}). 
This approach validates, although {\it a posteriori}, the use of the 
ef\/fective noise variance as representing the mean contribution of the 
instrumental noise and our previous results.
The CMB variances we obtain for the {\it Q}, {\it V}, and {\it W} coadded maps are 
$\sigma_{\mbox{\sc cmb}}^{2} = 5.98,\, 6.56,\, 6.77 \times 10^3 \mu\mbox{\rm K}^2$, 
respectively. 

Summarizing, the temperature distribution analyses of the coadded {\it Q}, {\it V}, 
and {\it W} CMB maps, as well as the eight individual DAs maps shown in 
fig.~\ref{figure1}, 
exibit that the distribution of the CMB temperature f\/luctuations does not 
obey a Gaussian distribution. 
The analysis of the other coadded CMB maps, mentioned before, produces a result 
that is fully consistent with this one.

Another way to test Gaussianity in a given data set is the calculation
of the factor involving the 4th- and 2nd- momenta of the distribution,
$R \equiv \langle {\Delta T}^{\,4} \rangle / {\langle 3 \Delta T^{\,2}\rangle}^2$, 
where $\langle x^n \rangle \equiv  \int \, \upd x \, P(x) \, x^n$  
($R=1$ corresponds to the Gaussian distribution case).
All the CMB maps analysed results in $R \in [1.02,1.06]$, which validates the 
possible description, within the four available ordinate decades, of the WMAP 
data using the $P_q$ distributions emerging from nonextensive statistical 
mechanics.

We also studied the coadded {\it W} map, after applying the Kp0 mask, in 
the four quadrants: North-East (NE), North-West (NW), South-West (SW), and 
South-East (SE) of the celestial sphere.
It turns out that the temperature distribution of the CMB data is sensibly  
different in one of these four sky patches. 
Indeed, we found $q_{_{\mbox{\sc ne}}} \simeq 1.015, 
\,q_{_{\mbox{\sc nw}}} \simeq 1.01$, $q_{_{\mbox{\sc sw}}} \simeq 1.02$, 
and $q_{_{\mbox{\sc se}}} \simeq 1.05$, with 
$R=1.02, 1.01, 1.025$, and $1.04$, respectively.

The significance of the $q$-index values found are based on the robustness 
of our findings. 
In fact, all our results are the same independently from the frequency-map 
or the coadded-map analysed. 
Also, they are equal for the two cut-sky masks applied, and for the 
one-year~\cite{BTV} and three-year maps.

\section{Monte Carlo analysis}
Now we present the results of our analyses considering the WMAP Monte Carlo 
(MC) simulations, which contain the instrument noise having a spatially 
varying noise structure, like real data. 

As a matter of fact, the WMAP detectors noises are inhomogeneous. 
This is due to the fact that the regions around the ecliptic poles (NW 
and SE quadrants) have 
large number-of-observations ($N_{\mbox{\rm\small obs}}$), and therefore 
low statistical noise, while the region corresponding to the ecliptic plane 
(NE and SW quadrants) were less observed (lower $N_{\mbox{\rm\small obs}}$ 
values), and have large statistical noise.
The spatially varying noise structure for each DA CMB map, with simulated 
instrument noise that include all known radiometric effects, were properly 
realized by the WMAP team in a set of MC sky realizations for each DA 
(available at 
{\mbox http:$\backslash\backslash$lambda.gsfc.nasa.gov$\backslash$product$\backslash\!\!$
map$\backslash$dr1$\backslash$noise$_{-}$sims$_{-}$imaps.cfm}). 
These MC data constitute an important tool for instrument noise evaluation 
corresponding to each DA, and also for the MC-coadded maps, obtainable 
for each one of the three bands ({\it Q}, {\it V}, and {\it W}) by combining 
MCs from different DAs.  
For instance, to obtain a MC-coadded CMB map in the {\it Q} band, one chooses 
{\it Q}1$^{i}$ from the set of MCs for the DA-{\it Q}1, and combine (in a coadded 
way, according to~\cite{Jarosik}) with {\it Q}2$^{j}$ from the set of MCs for the 
DA-{\it Q}2.

This time, the noise instrument of a given map is treated by computing the 
variance-normalized temperature for each pixel (a procedure described 
in~\cite{Spergel}: see, in particular, Eq. (17)). 
We observe that the instrument noise shows non-Gaussian behavior whose 
strength and signature are different to what is observed in our analyses 
of the 3-year WMAP data. 
In average, we obtained $\chi^2$ best-fits with distributions $P_q$ with 
$q=1.1 \pm 0.02$, and after the variance-normalization, the $q$-index 
maintain its value, although the variance is clearly modified.

\noindent
Moreover, we noticed that the non-Gaussian features of the instrument noise 
are sensible to $N_{\mbox{\rm\small obs}}$, in the sense that the $q$-index of 
the $P_q$ distributions changes significantly when analysing two regions in 
the sky, namely, one that was observed (at least) twice the other.
This interesting (and expected) feature is absent in the three-year 
WMAP data when compared with the (less observed) one-year WMAP 
data~\cite{BTV}, in fact we obtained essentially the same $q$-index for 
both data sets.
Finally, we have also to mention that such strong non-Gaussianities, present 
in all the coadded-MCs, exhibit a high symmetry between the SE and NW 
quadrants: $q_{_{\mbox{\sc se}}}/q_{_{\mbox{\sc nw}}} = 1.0 \pm 10^{-3}$. 
Again, this feature is fully absent in WMAP data, where we have  
$q_{_{\mbox{\sc se}}}/q_{_{\mbox{\sc nw}}} \simeq 1.04$.

\section{Comments and Conclusions}
Some comments are in due regarding the use of DA maps and the application of 
two WMAP masks, Kp0 and Kp2, in our analysis.
First, we analysed frequency by frequency ({\it Q}, {\it V}, and {\it W} bands) 
because the CMB Galactic foregrounds (free-free, synchrotron, and dust emission) 
are frequency dependent, which means that, if they are still present in the WMAP 
data, they should manifest differently at each frequency. 
Moreover, as recommended by the WMAP team, we have analysed the foreground 
reduced CMB maps~\cite{Lambda}, which were produced using the Foreground 
Template Model to subtract the synchrotron, free-free, and dust emission from 
the measured sky maps~\cite{WMAP3y}.
Our analyses of the one-point distributions of all these WMAP CMB maps do 
not show significant differences in the value of the $q$-index for the 
data investigated. 
Second, another way to examine the possible presence of unremoved foregrounds
is by analysing the CMB data using different cut-sky masks. 
We performed such analyses using both the Kp0 and Kp2 masks and, according to 
the $\chi^2$ best-fit estimator, we do not found significant difference in 
the values of the $q$-index.  
Third, known Galactic foregrounds contribute with positive temperature
f\/luctuations contaminations. 
Hence, we consider only the negative temperature f\/luctuations of the CMB 
maps to perform our data analyses (see the bottom plots of fig.~\ref{figure1}). 
These procedures, and their corresponding results, show that possible 
unremoved Galactic foregrounds could not be responsible for the detected 
non-Gaussian signatures. 

Now consider the instrument noise, which is always convolved with the CMB 
signal. 
In WMAP maps the instrument noise is inhomogeneous, higher in the Ecliptic 
Plane (roughly NE and SW quadrants) than near the Ecliptic Poles (NW and SE 
quadrants), and their contribution to $T_{\mbox{\rm\small pixel}}$ 
may cause a non-Gaussian behavior, as one can observe in MC sky maps 
simulated by the WMAP team. 
However, two interesting facts make the non-Gaussian noise signals 
qualitative and quantitatively different from what is observed in WMAP data. 
We emphasize that, in order to deal with inhomogeneous instrument noise 
in our CMB analyses, we have taked into account the following two procedures: 
the coadded sum of CMB DAs maps, as described in~\cite{Jarosik}, and the 
variance-normalized temperature, as described in~\cite{Spergel}. 
First, the instrument noise is extremely sensible to $N_{\mbox{\rm\small obs}}$, 
as expected, and as MC simulations confirm. 
It is well known that sky pixels in three-year WMAP data has been observed 
roughly 3 times more than one-year data, nevertheless the $P_q$ distributions 
are the best-fit one-point dsitributions, with $q=1.04 \pm 0.01$, for both 
data sets (see~\cite{BTV}). 
Second, the non-Gaussian signals from inhomogeneous instrument noise exibit 
a high symmetry between the NW and SE quadrants that has been quantifiable 
using the MC maps: 
$q_{_{\mbox{\sc se}}}/q_{_{\mbox{\sc nw}}} = 1.0 \pm 10^{-3}$, 
nonetheless this feature is fully absent in WMAP data, where we have  
$q_{_{\mbox{\sc se}}}/q_{_{\mbox{\sc nw}}} \simeq 1.04$.
These two facts strongly indicate that any subtle effect coming from 
inhomogeneous instrument noise contamination could not explain the presence 
of non-Gaussian signatures in the analysed WMAP data.

Summarizing, we have shown that a Gaussian distribution is excluded, at 
the 99\% CL, to properly represent the overall CMB temperature f\/luctuations 
measured by WMAP.
Through the temperature distribution analyses we notice that the three-year 
WMAP data are suitably explained, along about four decades, by a $P_q$ 
distribution with $q=1.04 \pm 0.01$. 
Moreover, our results also evidence that the strongest non-Gaussian 
contribution comes from the South-East quadrant, where  
$q_{_{\mbox{\sc se}}} \simeq 1.05$.
Therefore, the hypothesis of a nonextensive nature for the CMB temperature 
f\/luctuations is quite plausible.

\acknowledgments
\noindent
We acknowledge use of the Legacy Archive for Microwave Background Data
Analysis~\cite{Lambda} and of the HEALPix package~\cite{Gorski}.
A.B. acknowledges a PCI/DTI-MCT fellowship. 
C.T. acknowledges the partial support given by Pronex/MCT, CNPq and FAPERJ
(Brazilian Agencies).
T.V. acknowledges CNPq grant 305219/2004-9-FA.

\begin{largetable}
\caption{\rm All the variances are in units: $\times 10^3 \mu \mbox{\rm K}^2$.
Analyses of the 8 DA CMB maps, $N_{\mbox{\footnotesize\rm Kp0}} = 2,\/407,\/737$ 
pixels. 
DE means `direct evaluation' from the corresponding CMB map.
According to the $\chi^2$ best-fit estimator, the temperature distributions 
of these CMB data are described by $P_q$ distributions with variance 
$\sigma_{\nu}^{2}$ and $q$-index: $q=1.04 \pm 0.01$.}
\vspace{0.3cm}
\label{table1}
\begin{center}
\begin{tabular}{cccccccc}  
\hline \\
DA map 
& $\sum(1/n_i)$ 
& ${\sigma_{0}}^{2}$ 
& $\sigma_{\mbox{\rm\small noise}}^{2}$ & $\sigma_{\nu}^{2}$
& $\sigma_{_{\mbox{\footnotesize\sc cmb}}}^{2}=\sigma_{\nu}^{2} 
- \sigma_{\mbox{\rm\small noise}}^{2}$  
& $\sigma_{\nu \mbox{\footnotesize\sc -de}}^{2}$ 
& $\sigma_{_{\mbox{\footnotesize\sc cmb-de}}}^{2}
\!=\sigma_{\nu \mbox{\footnotesize\sc -de}}^{2} 
- \sigma_{\mbox{\rm\small noise}}^{2}$ 
\vspace{0.3cm} \\ 
\hline \\
{\it Q}1  &  4925.6  &  5039.58   &  10.31  &  16.03  &  5.7  &  17.02  &  6.7  \\
{\it Q}2  &  4921.9  &  4556.94   &   9.32  &  14.93  &  5.6  &  15.94  &  6.6  \\
{\it V}1  &  3655.2  &  10916.4   &  16.57  &  22.94  &  6.4  &  23.93  &  7.4  \\
{\it V}2  &  3655.3  &  8677.74   &  13.17  &  19.46  &  6.3  &  20.60  &  7.4  \\
{\it W}1  &  2437.1  &  34613.2   &  35.04  &  40.32  &  5.3  &  42.81  &  7.8  \\
{\it W}2  &  2444.8  &  42672.3   &  43.33  &  48.08  &  4.8  &  50.72  &  7.4  \\
{\it W}3  &  2445.0  &  47401.8   &  48.14  &  52.63  &  4.5  &  55.53  &  7.4  \\
{\it W}4  &  2430.4  &  45482.9   &  45.91  &  50.51  &  4.6  &  53.71  &  7.8  \\
\hline
\end{tabular}
\end{center}
\end{largetable}

\end{document}